\documentclass[letter]{aa} 

\usepackage{graphicx}
\usepackage{txfonts}
\usepackage{hyperref}  
\hypersetup{colorlinks=true,linkcolor=[rgb]{1.,0.2,0.2},citecolor=[rgb]{0.1,0.4,1.},filecolor=[rgb]{0.7,0.2,0.2},urlcolor=[rgb]{0.7,0.2,0.2}}

\definecolor{blue}{rgb}{0., 0., 1}

\newcommand{\heialone}{\textrm{He}\textsc{i}}

\newcommand{\oii}{[\textrm{O}\textsc{ii}]}
\newcommand{\oiidoub}{[\textrm{O}\textsc{ii}]}
\newcommand{\oiidoublam}{[\textrm{O}\textsc{ii}]\ensuremath{\lambda3727,3729}}

\newcommand{\nevalone}{[\textrm{Ne}\textsc{v}]}
\newcommand{\nev}{[\textrm{Ne}\textsc{v}]\ensuremath{\lambda3426}}
\newcommand{\neiii}{[\textrm{Ne}\textsc{iii}]}
\newcommand{\neiiibluelam}{[\textrm{Ne}\textsc{iii}]\ensuremath{\lambda3869}}
\newcommand{\neiiiredlam}{[\textrm{Ne}\textsc{iii}]\ensuremath{\lambda3967}}

\newcommand{\oiiiaur}{[\textrm{O}\textsc{iii}]\ensuremath{\lambda4363}}
\newcommand{\oiiiv}{[\textrm{O}\textsc{iii}]\ensuremath{\lambda5007}}

\newcommand{\ha}{\ifmmode {\rm H}\alpha \else H$\alpha$\fi}
\newcommand{\halam}{\ifmmode {\rm H}\alpha \lambda6563 \else H$\alpha$ $\lambda$6563 \fi}
\newcommand{\hb}{\ifmmode {\rm H}\beta \else H$\beta$\fi}
\newcommand{\hblam}{\ifmmode {\rm H}\beta \lambda4861 \else H$\beta$ $\lambda$4861 \fi}
\newcommand{\hg}{\ifmmode {\rm H}\gamma \else H$\gamma$\fi}
\newcommand{\hglam}{\ifmmode {\rm H}\gamma \lambda4340 \else H$\gamma$ $\lambda$4340 \fi}
\newcommand{\hd}{\ifmmode {\rm H}\delta \else H$\delta$\fi}
\newcommand{\hdlam}{\ifmmode {\rm H}\delta \lambda4102 \else H$\delta$ $\lambda$4102 \fi}
\newcommand{\lya}{\ifmmode {\rm Ly}\alpha \else Ly$\alpha$\fi}
\newcommand{\pg}{\ifmmode {\rm P}\gamma \else Pa$\gamma$\fi}
\newcommand{\lyb}{\ifmmode {\rm Ly}\beta \else Ly$\beta$\fi}
\newcommand{\lyg}{\ifmmode {\rm Ly}\gamma \else Ly$\gamma$\fi}

\newcommand{\niv}{\textrm{N}\textsc{iv}]\ensuremath{\lambda1486}}

\newcommand{\civ}{\textrm{C}\textsc{iv}\ensuremath{\lambda1548,1550}}
\newcommand{\civblue}{\textrm{C}\textsc{iv}\ensuremath{\lambda 1548}}

\newcommand{\flyc}{\ifmmode  \mathrm{f}_\mathrm{esc}\mathrm{(LyC)} \else $\mathrm{f}_\mathrm{esc}\mathrm{(LyC)}$\fi}

\def\sfr{M$_{\odot}$~yr$^{-1}$}
\def\sfrsd{M$_{\odot}$~yr$^{-1}$~kpc$^{-2}$}

\def\ergs{\ifmmode \mathrm{erg\hspace{1mm}s}^{-1} \else erg s$^{-1}$\fi}
\def\ergscm{erg s$^{-1}$ cm$^{-2}$}
\def\micron{\ifmmode \mu\mathrm{m} \else $\mu$m\fi}
\def\msun{\ifmmode \mathrm{M}_{\odot} \else M$_{\odot}$\fi}
\def\msunyr{\ifmmode \mathrm{M}_{\odot} \hspace{1mm}{\rm yr}^{-1} \else $\mathrm{M}_{\odot}$ yr$^{-1}$\fi}
\def\zsun{\ifmmode Z_{\odot} \else Z$_{\odot}$\fi}
\def\lsun{\ifmmode L_{\odot} \else L$_{\odot}$\fi}
\def\mstar{\ifmmode \mathrm{M}_{\star} \else M$_{\star}$\fi}
\newcommand{\JWST}{\textrm{JWST}}

\usepackage{orcidlink}

\begin{document}

\titlerunning{\JWST\ probes massive galaxies $z\simeq11$}
\title{DREAMS: Taking the Temperature of a Massive Galaxy at $z \simeq11$}
\authorrunning{Lorenzo Parente et al.}
\author{
L.~Parente\inst{\ref{unibo}}\fnmsep\thanks{E-mail: \href{mailto:lorenzo.parente@studio.unibo.it}{lorenzo.parente@studio.unibo.it}}$^{\orcidlink{0009-0000-0040-9138}}$ \and
E.~Vanzella\inst{\ref{inafbo}}$^{\orcidlink{0000-0002-5057-135X}}$ \and
M.~Messa\inst{\ref{inafbo}}$^{\orcidlink{0000-0003-1427-2456}}$ \and
A.~Zanella\inst{\ref{inafbo}}$^{\orcidlink{0000-0001-8600-7008}}$\and
K.~Nakajima\inst{\ref{Kanazawa},\ref{naoj}}$^{\orcidlink{0000-0003-2965-5070}}$\and
M.~Ouchi\inst{\ref{naoj},\ref{icrr},\ref{sokendai},\ref{ipmu}}$^{\orcidlink{0000-0002-1049-6658}}$\and
R.~A.~Windhorst\inst{\ref{asu}}$^{\orcidlink{0000-0001-8156-6281}}$\and
M.~Talia\inst{\ref{unibo},\ref{inafbo}}$^{\orcidlink{0000-0003-4352-2063}}$\and
Y.~Harikane\inst{\ref{icrr}}$^{\orcidlink{0000-0002-6047-430X}}$\and
Y.~Ono\inst{\ref{icrr}}$^{\orcidlink{0000-0001-9011-7605}}$\and
A.~M.~Koekemoer\inst{\ref{stsci}}$^{\orcidlink{0000-0002-6610-2048}}$\and
P.~Bergamini\inst{\ref{inafbo}}$^{\orcidlink{0000-0003-1383-9414}}$ \and
M.~Meneghetti\inst{\ref{inafbo}}$^{\orcidlink{0000-0003-1225-7084}}$ \and
P.~Rosati\inst{\ref{unife},\ref{inafbo}}$^{\orcidlink{0000-0002-6813-0632}}$ \and
Y.~Xu\inst{\ref{dawn},\ref{naoj}}$^{\orcidlink{0000-0002-5768-8235}}$ \and
H.~Umeda~\inst{\ref{Tsukuba},\ref{icrr},\ref{sokendai}}$^{\orcidlink{0009-0008-0167-5129}}$ \and
Y.~Zhang\inst{\ref{Pasadena}}$^{\orcidlink{0000-0003-3817-8739}}$ \and
Y.~Isobe\inst{\ref{KavliUK}}$^{\orcidlink{0000-0001-7730-8634}}$
}
\institute{
Dipartimento di Fisica e Astronomia, Università degli Studi di Bologna, Via Gobetti 93/2, 40129 Bologna, Italy \label{unibo}
\and
INAF -- OAS, Osservatorio di Astrofisica e Scienza dello Spazio di Bologna, via Gobetti 93/3, I-40129 Bologna, Italy \label{inafbo} 
\and
Institute of Liberal Arts and Science, Kanazawa University, Kanazawa, Japan 
\label{Kanazawa}
\and
{National Astronomical Observatory of Japan, 2-21-1 Osawa, Mitaka, Tokyo 181-8588, Japan} \label{naoj}
\and
{Institute for Cosmic Ray Research, The University of Tokyo, 5-1-5 Kashiwanoha, Kashiwa, Chiba 277-8582, Japan} \label{icrr}
\and
{Astronomical Science Program, Graduate Institute for Advanced Studies, SOKENDAI, 2-21-1 Osawa, Mitaka, Tokyo, Japan} \label{sokendai}
\and
{Kavli Institute for the Physics and Mathematics of the Universe (WPI), The University of Tokyo, 5-1-5 Kashiwanoha, Kashiwa, Chiba 277-8583, Japan} \label{ipmu}
\and
{School of Earth and Space Exploration, Arizona State University, Tempe, AZ 85287-6004, USA} \label{asu}
\and
{Space Telescope Science Institute, 3700 San Martin Drive, Baltimore, MD 21218, USA} \label{stsci} 
\and
Dipartimento di Fisica e Scienze della Terra, Università degli Studi di Ferrara, Via Saragat 1, I-44122 Ferrara, Italy\label{unife}
\and
Cosmic Dawn Center (DAWN), Niels Bohr Institute, University of Copenhagen, Jagtvej 128, DK-2200 Copenhagen N, Denmark\label{dawn}
\and
Center for Computational Sciences, University of Tsukuba, Ten-nodai, 1-1-1 Tsukuba, Ibaraki 305-8577, Japan\label{Tsukuba}
\and
IPAC, California Institute of Technology, MC 314-6, 1200 E. California Boulevard, Pasadena, 91125, CA, USA\label{Pasadena}
\and
Kavli Institute for Cosmology, University of Cambridge, Madingley Road, Cambridge, CB3 0HA, UK\label{KavliUK}
}

\date{} 

 
\abstract 
{
We present a detailed photometric and spectroscopic analysis of Janus-11, a $z=10.7933 \pm 0.0006 $ galaxy magnified by a factor 1.6 by the galaxy cluster MACS J0416. It was observed with JWST/NIRCam as part of the PEARLS program and with JWST/NIRSpec MSA as part of the DREAMS program. It has M$_{\rm UV}=-19.78^{+0.12}_{-0.11}$, and its stellar mass of $\log(\mstar/\msun)=8.75^{+0.08}_{-0.10}$ places it among the most massive known $z>10$ galaxies, while its $A_V=0.54\pm0.09$ mag and UV slope of $\beta=-1.71\pm0.26$ are compatible with a red monster classification, albeit among the least obscured members of this population: this could hint at a transition from the red monster to the blue monster phase. Moreover, the \oiiiaur\ detection allows for the highest-redshift metallicity measurement based on the \oiiiaur\ direct method to date of $12 + \log(O/H) = 7.74^{+0.11}_{-0.10}$, together with an electron temperature of $T_e$(O$^{++}$) = $(2.1 \pm 0.2)\times10^4$ K. Its spatially resolved nature, with an effective radius of $136 \pm 9$\,pc, together with the narrow FWHM of the emission lines ($\lesssim 230$~km~s$^{-1}$, or $\sigma_v \lesssim 100$~km~s$^{-1}$), and the lack of significant detections of high-ionization emission lines and broad components, favor an SFG classification, although an AGN contribution cannot be excluded.
Janus-11 joins the small sample of spectroscopically confirmed and characterized sources at $z>10$, providing new information on massive systems with non-negligible dust attenuation and well-measured metallicity in this epoch.
}
   \maketitle

\section{Introduction}
\label{sect:intro}

Determining interstellar medium (ISM) and stellar properties of high-redshift galaxies is of crucial importance to test physical models of galaxy formation and evolution, but also to extend low-redshift relations. With the advent of JWST, it has become possible to obtain spectra with low to high spectral resolution, with multiple emission-line detections up to redshift z$\sim$14 \citep{ArrabalHaro2023, Carniani2024, Naidu2026}. This has enabled measurements of important properties such as ionization conditions, metallicities and star formation rates (SFR) in real high-redshift environments, whereas previously such measurements were limited to high-redshift analogues \citep{Laseter2022, Thuan2022}. However, there are still only a few well-characterized objects at $z>10$ (\citealp{Larson2026}, \citealp{Ferrara2026}) and their physical conditions vary widely, leading, for example, to high N/O or C/O abundances \citep{DEugenio2024, Topping2024b}. Also, the observed number density of UV luminous galaxies is higher than expected, in tension with theoretical models \citep{Finkelstein2024, Harikane2025}. While there are multiple possible explanations for these peculiarities, none of them can be conclusively ruled in or out given the current data \citep{Cueto2026}.
For these reasons, it is of fundamental importance to increase the number of spectroscopically characterized sources at $z>9.5-10$, where optical emission lines such as \hb~and \oiiiv~are redshifted out of the NIRSpec wavelength range.

High-redshift galaxies tend to have low metallicities, high electron densities and high ionization \citep{Curti2024, nakajima2023}. In particular, their metallicity is related to processes such as the cooling of ISM and IGM \citep{Smith2008} and the production of ionizing photons \citep{Ciardi2003} and dust \citep{Draine2003}. Moreover, it is connected to the baryon cycle and in particular to the SFR and stellar mass by a relation known as Fundamental Metallicity Relation \citep[FMR,][]{mannucci10}, which is well characterized up to z<3.3 \citep{Sanders2021}. The main problem in measuring the metallicity is the low intensity of the \oiiiaur~auroral emission line, which allows the electron temperatures and then the metallicity to be estimated through the robust "direct method". Most studies thus rely on the strong line method \citep[e.g.,][]{Curti2017}, which exploits optical nebular lines; however, calibrations of this method are based on low-redshift samples and have been shown to be unreliable at high redshift \citep{Laseter2024}. While recent studies have attempted to extend these calibrations to higher redshifts \citep{Sanders2024, Sanders2026}, the scarcity of robust direct method measurements remains a major limitation not only for the calibrations themselves, but also for obtaining accurate metallicity estimates and establishing their connection with the physical properties of galaxies \citep{Nishigaki2025, Perez2026}.

In this work, we present a detailed photometric and spectroscopic analysis of a $z=10.793$ source with detected \oiiiaur, observed with both JWST/NIRCam and NIRSpec. This is the highest-redshift \oiiiaur-based direct-method metallicity measurement to date. This galaxy is located in the MACS J0416 field, which weakly lenses the source, and was first discovered by \citet{CanucsDR1_2026}.
We assume a flat $\Lambda$CDM cosmology with $\Omega_{\rm M}=0.315$ and $H_{0}=67.4\,{\rm km}\,{\rm s}^{-1}\,{\rm Mpc}^{-1}$ \citep{Planck2020}, giving a scale of 4.026 kpc/" and age of 423 Myr at z=10.793. Magnitudes are in the AB system \citep{Oke_1983}, $m_{\rm AB}=23.9-2.5\log(f_\nu/\mu{\rm Jy})$.

\begin{figure*}
\center
 \includegraphics[width=1.99\columnwidth]{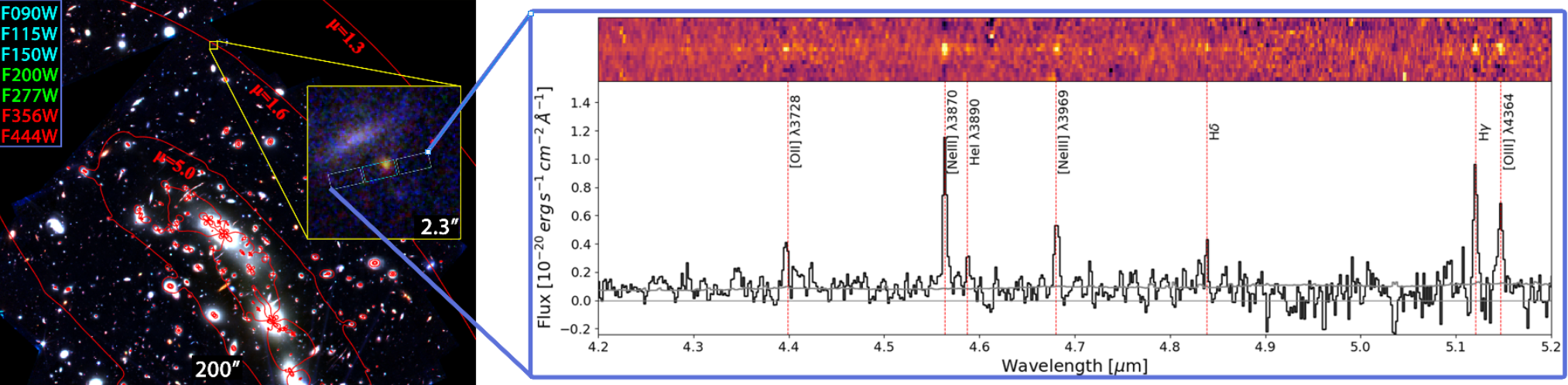}
 \caption{Left: NIRCam color image of the MACS J0416 cluster (blue: F090W, F115W, F150W; green: F200W, F277W; red: F356W, F444W). The red contours show $\mu=5.0, 1.6, 1.3$ at $z=10.8$. The inset shows Janus-11 and the foreground galaxy with MSA shutters superimposed. Right: NIRSpec LW spectrum from 4 to 5.2\,$\mu m$; the grey line represents the 1$\sigma$ error.} 
 \label{fig:overview}
\end{figure*}

\begin{figure}
\center
 \includegraphics[width=0.49\textwidth]{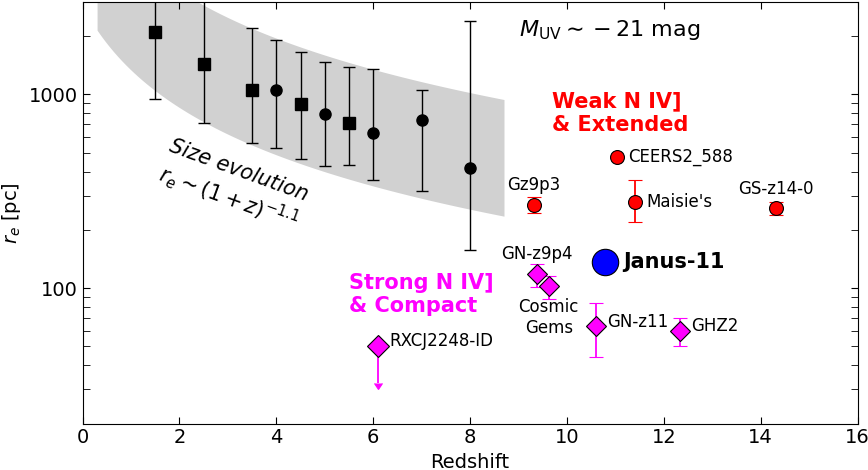}
 \caption{Adapted from \citet{Harikane2025}. Rest frame UV effective radius plotted against redshift. Black symbols represent the size evolution of bright galaxies ($M_{\rm UV}\sim-21$) at z=0--8 with 1$\sigma$ dispersion \citep{Shibuya2015}, red symbols represent z>9 galaxies following the z=0--8 relation and magenta symbols represent compact galaxies with $r_e<100\,pc$. Janus-11 is represented as a blue circle.
 }
 \label{fig:size_redshift}
\end{figure}
%


\section{\JWST\ observations and analysis} \label{sec:data}
Our analysis is based on JWST/NIRCam images from the Prime Extragalactic Areas for Reionization and Lensing Science program (PEARLS, program 1176, PI: R. Windhorst; \citealt{windhorst23_pearls}), whose mosaics also incorporate data from other public archival programs, as detailed by \citet{Bolamperti2026}.

We use JWST/NIRSpec MSA data from the program GO-4750 (PI: K. Nakajima), titled Deep Reconnaissance of Early Assemblies of Metal-poor Star formation (DREAMS, Nakajima et al. in prep), targeting the field of the lensing cluster MACS J0416. It provides spectroscopic data acquired with the medium-resolution gratings G140M/F070LP and G395M/F290LP. 

Being at the edge of the field, Janus-11 is magnified by a factor $\mu = 1.6$ \citep{Bergamini2023b}. The statistical uncertainty on this value is $<1\%$, and is thus negligible.
Figure~\ref{fig:overview} shows the NIRCam colour image of Janus-11 and the surrounding field, together with the NIRSpec long-wavelength (LW) spectrum.

We perform aperture photometry using a circular aperture to measure the object flux and an annulus to estimate the background, both centered on Janus-11. Specifically, we perform a Monte Carlo simulation perturbing the pixel values in both the circular aperture and the background annulus by drawing from a Gaussian distribution centered on the image pixel value and with standard deviation derived from the weight maps. We then apply an aperture correction using PSFs computed with STPSF \citep{Perrin2014}, and derive the fluxes and uncertainties as the median and the 16th/84th percentiles of the distributions (Table~\ref{tab:photometry}).

Line fitting is performed using lmfit \citep{lmfit1.3.4} and emcee \citep{emcee_2013}. Each line is first fitted with lmfit and then the resulting best-fit parameters are used as initial guesses for the emcee MCMC. \niv, \nev, \oiidoub, \neiiibluelam, \neiiiredlam, \heialone\,$\lambda3889$ and \hd~are fitted independently from each other with a composite model of a Gaussian and a zero-order polynomial with all free parameters; \hg~and \oiiiaur~are fitted simultaneously with a two-Gaussian and zero-order polynomial model, in order to better estimate the continuum given their proximity. Also, the line width of \oiiiaur\ is fixed to the width of \hg\, as it presents a blue tail that we consider as an effect of noise, since it is not observed in \hg\ and \neiiibluelam\ lines.
Measured line properties are reported in Table~\ref{tab:lines}. Except for the high ionization emission lines, all lines are detected with an $\textrm{SNR} > 3$. All detected lines are also spectrally resolved, with FWHMs larger than the instrumental resolution FWHM of $\sim$39.4\,\AA~\citep{jakobsen2022}; the \hd\ line is the only exception, as it is compatible with being unresolved. Nonetheless, they appear as narrow lines, with $\textrm{FWHM}\lesssim 230$~km~s$^{-1}$, or $\sigma_v \lesssim 100$~km~s$^{-1}$.

All quantities discussed below are reported in Table~\ref{tab:properties}.

\begin{table}
   \caption{Derived quantities.} 
   \label{tab:properties}      
   \centering     
   \setlength{\tabcolsep}{5pt}     
   \begin{tabular}{l c }   
      \hline\hline  
      Quantity    & Value  \\ 
      \hline
      $\mu_{\rm tot}$,~$\mu_{\rm tang}$ &  $1.60, 1.57$ \\
      M$_{\rm UV}^{(\ast)}$ & $-19.78^{+0.12}_{-0.11}$ \\
      log(\mstar/\msun)$^{(\ast)}$ & $8.75^{+0.08}_{-0.10}$ \\
      $A_V$ & $0.54 \pm 0.09$ \\
      SFR$_{10}^{(\ast)}$ [\sfr] & $9^{+3}_{-2}$ \\
      $\beta$ & -1.71±0.26 \\
      $r_e$$^{(\S)}$$^{(\ast)}$ [pc] & $136 \pm 9$ \\
      \hline
      \ha$^{(\ast)}$\ ~~[$10^{-18}$ erg~s$^{-1}$~cm$^{-2}$] &  $4.0^{+0.8}_{-0.7}$ \\
      \hb$^{(\ast)}$\ ~~[$10^{-18}$ erg~s$^{-1}$~cm$^{-2}$] &  $1.5^{+0.3}_{-0.2}$ \\
      \oiiiv$^{(\ast)}$\ ~~[$10^{-17}$ erg~s$^{-1}$~cm$^{-2}$] &  $1.4^{+0.3}_{-0.2}$ \\
      \hline
      \hg/\hd &  $3.5^{+1.3}_{-0.9}$ \\
      log(Ne3O2) &  $0.43^{+0.12}_{-0.10}$ \\
      log(\oiiiaur/\hg) &  $-0.14 \pm 0.07$ \\
      O32&  $46^{+16}_{-10}$\\
      \oiiiv/\oiiiaur &  $28^{+5}_{-4}$ \\
      \hline
      SFR$_{\textrm\hg}^{(\ast)}$ [\sfr] &  $20\pm4$ \\
      sSFR [$Gyr^{-1}$] & $36\pm10$ \\
      $\Sigma_{SFR}$ [\sfrsd] & $172\pm41$ \\
      $T_e$(O$^{++}$) [$10^4$ K] &  $2.1 \pm 0.2$ \\
      Z(Z$_{\odot}$), 12+log(O/H)$^{\ddagger}$ &  $0.11^{+0.03}_{-0.02}$, $7.74^{+0.11}_{-0.10}$ \\
      log U &  $-1.69^{+0.10}_{-0.09}$ \\
      \hline \hline
   \end{tabular}
   \tablefoot{  
   Errors are at $1\sigma$ and do not include systematics due to the scatter in the relations from \citet{Witstok2021} and \citet{Izotov2006}. 
   $^{(\S)}$ $r_e$ [pc] = $a$ [pix] $\cdot$ $\sqrt{q}$ $\cdot$ masPerPixel $\cdot$ scale / $\sqrt{\mu_{\rm tot}}$, where $a$ [pix] = $2.5\pm0.15$ is the semi-major axis, $q=b/a=0.73\pm0.05$ is the axis ratio, $1\,\mathrm{pix}=20\,\mathrm{mas}$, and the scale is $4026\,\mathrm{pc}/\arcsec$.
   $^{(\ast)}$ Corrected for magnification.
   $^{(\ddagger)}$ From \citet{asplund2009}, $Z_{\odot} \Rightarrow 8.69=12+\log({\rm O/H})$.  
   }
\end{table}

\section{Results} \label{sec:discussion}
\subsection{SED fitting and morphology} \label{sec:sedAndMorph}
SED Fitting is performed using BAGPIPES (v1.3.6; \citealt{Carnall2018}). We use a stellar grid derived using BPASS v2.2.1 \citep{BPASS}, including binary stars, with a \citet{Kroupa_2001} IMF and nebular contribution computed via CLOUDY \citep{Gunasekera2025}.
We perform the SED fitting incorporating both the photometric SED and the LW spectrum. We assume a ``delayed'' star-formation history with log-uniform priors in the parameters age and tau, and a Calzetti extinction curve \citep{Calzetti2000}. From this fit, we find a stellar mass of log(\mstar/\msun)$=8.75^{+0.08}_{-0.10}$, a dust attenuation of $A_V=0.54\pm0.09$\,mag and an SFR in the last 10 Myr of $\textrm{SFR}_{10}=9^{+3}_{-2}$\,\sfr. Both the stellar mass and the $A_V$ are among the highest at $z>10$ (\citealp{Ferrara2026}). The SED fitting results are shown in Figure~\ref{fig:sed_calzetti}. An alternative fit assuming a Salim extinction curve, which also hints at the possible presence of the $2175\AA$ bump, is described in Appendix \ref{sec:appendix_sed}.

We measure the rest-frame UV continuum slope by fitting a power law ($F_\lambda\propto\lambda^\beta$) to the F200W, F277W and F356W filters, corresponding to a rest-frame wavelength range of 1488-3376 \AA, finding $\beta = -1.71 \pm 0.26$. This moderately red UV-slope, while not being unique, is unusual at this redshift, as recent studies find median values $< -2.1$, with single-galaxy UV-slopes often compatible with the dust-free limit of $\beta = -2.6$ \citep{Bolamperti2023, Topping2024a, Cullen2024}. Such red $\beta$ values may require a rapid build-up of dust or a significant contribution of nebular emission to the UV continuum \citep{Saxena2026}. Moreover, together with the $A_V$ value estimated by SED fitting, it hints at a red monster classification, although this is an intermediate value between other red monsters and blue monsters.

Janus-11 is spatially resolved in all filters and has a 
magnification-corrected circularized effective radius $r_e= 136 \pm 9\,\mathrm{pc}$. 
This object therefore lies midway between the compact and extended classifications of \citet{Harikane2025}, as shown in Figure~\ref{fig:size_redshift}. However, the lack of significant detections of high-ionization emission lines, such as \niv\ and \civ, 
more closely resemble the properties of the extended population. 
With an absolute magnitude in the F200W filter of $M_{\rm UV}$=$-19.78^{+0.12}_{-0.11}$, Janus-11 is fainter than the sample in \citet{Harikane2025}, which is limited to $-24<M_{\rm UV}<-21$.

\subsection{Lines-derived physical properties} \label{temp_logu_metal}
We derive the ionization parameter and the metallicity following \citet{Pascalau2026} and estimate the \oiiiv~from the \neiii/\oii~ratio using Eq.~1 of \citet{Witstok2021}.
While the \hg/\hd\ ratio is too uncertain to provide a good estimate of the nebular extinction and leads to unphysical values, we correct all measured line fluxes for dust extinction assuming $E(B-V)_{\rm neb}=E(B-V)_{\rm star}+k$ with $k=0.053\pm0.025$, adopting the relation found at $5<z<7$ by \citet{Karthikeyan2026}. We estimate the \ha~and \hb~fluxes from the \hg~flux adopting the theoretical Balmer decrements from the emissivities at the estimated Te(O$^{++}$) using PyNeb \citep{Luridiana2015}. Moreover, we assume an electron density $n_e$ of 1000\,cm$^{-3}$, after checking there are no significant variations in the derived quantities over the range 100--10000\,cm$^{-3}$. Lastly, to compute Te(O$^{+}$) we use the intermediate-Z relation from \citet{Izotov2006}.
We find Te(O$^{++}$)$=2.1\pm0.2\times10^4\,\mathrm{K}$, an ionization parameter of $\log~U=-1.69^{+0.10}_{-0.09}$ and a metallicity of $12 + \log(O/H) = 7.74^{+0.11}_{-0.10}$ or $Z=0.11^{+0.03}_{-0.02}\,\zsun$. This represents the highest-redshift \oiiiaur-based direct-method metallicity measurement, extending the record set by GN-z11 \citep{Alvarez-Marquez2025}. \newline
We also find compatible metallicity and log U values using the software HII-CHI-mistry (Appendix \ref{sec:HII_CHI_mistry}).

Adopting the calibration of \citet{Reddy2018}, the SFR inferred from the \ha\ luminosity (Sect.~\ref{temp_logu_metal}) is SFR$_\textrm{\hg}=20\pm4$~\sfr. The conversion factor of \citet{Reddy2018} is better suited to high-redshift galaxies, as it is derived from a model with a metallicity of 0.28~\zsun. We note that the SFR$_\textrm{\hg}$ is slightly higher than the photometric SFR$_{10}$.
From the stellar mass and the effective radius derived in Sect.~\ref{sec:sedAndMorph} we find a specific star formation rate and a star formation rate surface density of $36\pm10\,Gyr^{-1}$ and $\Sigma_{SFR}=\frac{SFR}{2\pi r_e^2}=172\pm41$\,\sfrsd, respectively.

\subsection{SFG or AGN}
\begin{figure}
\center
 \includegraphics[width=0.46\textwidth]{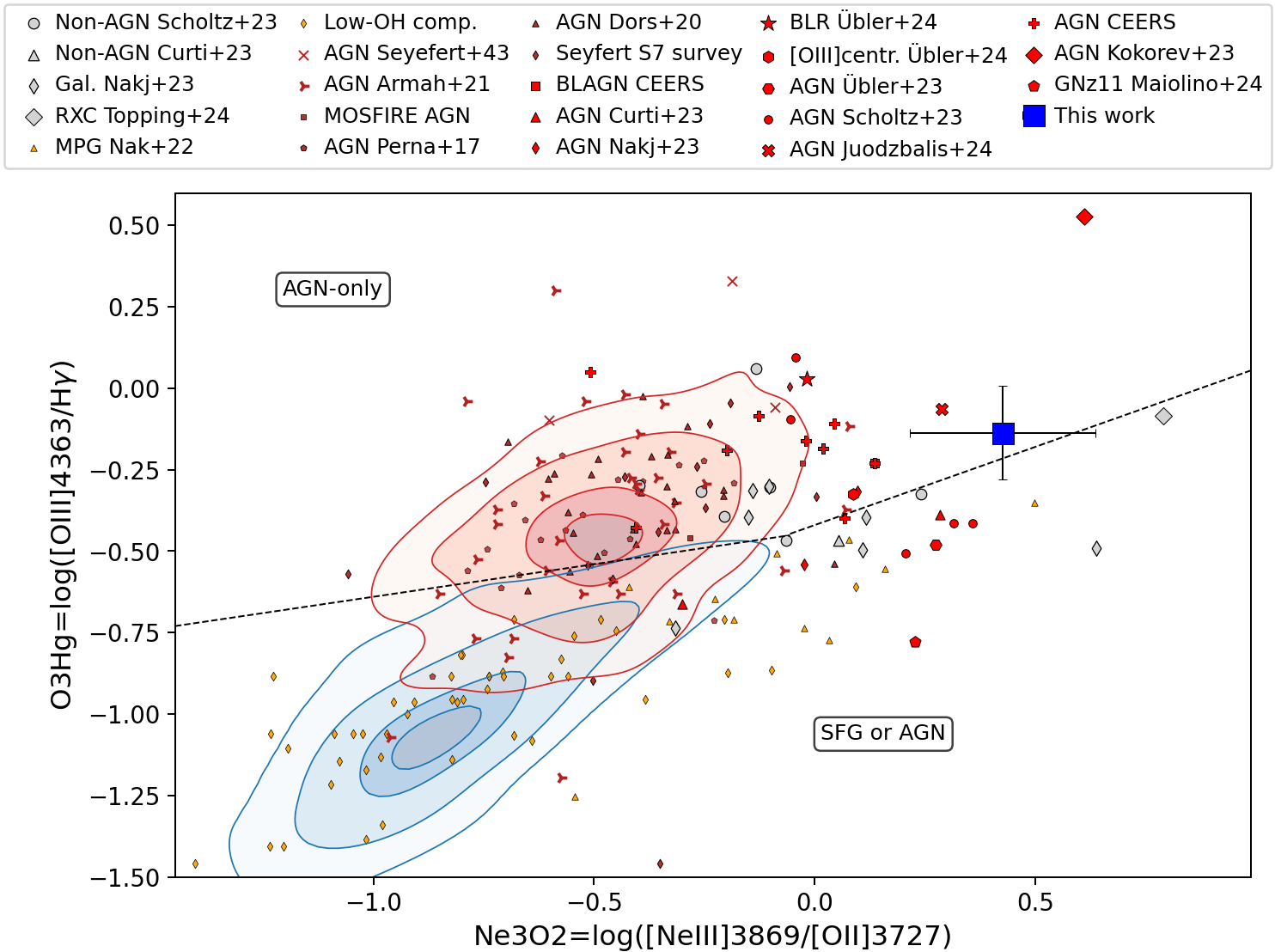}
 \caption{Diagnostic diagram adapted from \citet{Mazzolari2024}. The region upward the red-dashed line is AGN-only, while sources downward could be SFGs or AGNs. AGNs are represented in red, SFGs in orange and high-z sources not classified as AGN in grey. The red and blue contours show the distribution of AGN and SFGs from the SDSS, respectively. Janus-11 is reported as a blue square with 2$\sigma$ errors.
 }
 \label{fig:mazzolari_diagram}
\end{figure}

While the BPT diagnostic diagrams \citep{BPT1981} do not predict correctly the nature of very high-redshift extragalactic objects because of the low metallicities and extreme ionization conditions of the ISM, \citet{Mazzolari2024} proposed 
diagnostic diagrams based on different line ratios calibrated for high-redshift environments. One of these diagrams uses the $\log(\oiiiaur/\hg)$ and the Ne3O2 ratios, both derived from lines directly measured in our spectrum. Janus-11 falls in the AGN region (Figure \ref{fig:mazzolari_diagram}), but is compatible with the AGN and SFG mixed region within 2$\sigma$. 
From the morphological fit we find a 2$\sigma$ upper limit on a PSF total flux contribution of $7\%$ in the F200W filter, as described in Appendix~\ref{sec:appendix_morph}.
The absence of broad line components and high-ionization lines (such as \nevalone)
and the spatially-resolved morphology suggest a SFG classification, although a contribution from an AGN cannot be excluded.

\section{Final remarks}
Janus-11 at $z=10.7933 \pm 0.0006 $ is among the most massive sources discovered to date at $z>10$ ($\log(\mstar/\msun)=8.75^{+0.08}_{-0.10}$). With a dust attenuation of $A_V=0.54\pm0.09$ mag and a UV-slope of $\beta = -1.71 \pm 0.26$, it is compatible with a red monster classification, although it would be the least obscured in this population; as such, this object could be related to a transition from the red monster to the blue monster phase, as proposed in the AFM model by \citet{Ferrara2026}. Moreover, it presents a strong \oiiiaur~detection with S/N=8.5, which enables the highest-redshift metallicity measurement with the \oiiiaur~based direct method, resulting in 
$Z=0.11^{+0.03}_{-0.02}\,\zsun$, and an electron temperature measurement of $T_e$(O$^{++}$) = $(2.1 \pm 0.2)\times10^4$ K. The source appears to be spatially resolved with an effective radius of $136 \pm 9$\,pc, for which we estimate a SFR of $20\pm4$\,\sfr~derived from the \hg\,, which is slightly higher than the SFR inferred from the SED fitting of $9^{+3}_{-2}$\,\sfr.
Finally, the source lies on the dividing line between AGNs and SFGs in the diagnostic diagram of \citet{Mazzolari2024}. The lack of prominent broad spectral components and high-ionization lines, together with its spatially resolved nature, favors an SFG classification, although an AGN contribution cannot be entirely ruled out.



\begin{acknowledgements}
This work is based on observations made with the NASA/ESA/CSA 
\textit{James Webb Space Telescope} (\JWST).
These observations are associated with \JWST\ GO program n.4750 (DREAMS, PI K. Nakajima), GTO n.1176 (PEARLS, PI R. Windhorst). 
We thank G.Mazzolari for providing the data of Figure~\ref{fig:mazzolari_diagram}. 
EV and MM acknowledge financial support through INAF GO Grant 2024 ``Mapping Star Cluster Feedback in a Galaxy 450 Myr after the Big Bang'' and the project PRORIS - COSMOWEB ``A new era for cosmology:  exploiting the JWST revolution''.  
RAW acknowledges support from NASA JWST Interdisciplinary Scientist grants
NAG5-12460, NNX14AN10G and 80NSSC18K0200 from GSFC.
M.O. acknowledges the support from the World Premier International Research Center Initiative 
(WPI Initiative), MEXT, Japan, the joint research program of the Institute for Cosmic Ray Research (ICRR), 
the University of Tokyo, and KAKENHI (21H04467, 25H00674) through Japan Society 
for the Promotion of Science (JSPS).
PB acknowledges financial support from the Italian Space Agency (ASI) through contract ``Euclid - Phase E'', INAF Grants ``The Big-Data era of cluster lensing'' and ``Probing Dark Matter and Galaxy Formation in Galaxy Clusters through Strong Gravitational Lensing''.
\end{acknowledgements}

\bibliographystyle{aa}
\bibliography{bib}

\appendix
\onecolumn
\section{Measured photometric fluxes and line properties}

\begin{table}[htbp]
   \caption{Photometry.} 
   \label{tab:photometry}      
   \centering          
   \begin{tabular}{l c  }   
      \hline\hline  
      Quantity    & Value  \\ 
      \hline
      RA & 04:16:09.43\\
      Dec & $-$24:02:44.6\\
      Redshift$^{\dag}$ & $10.7933 \pm 0.0006 $\\
      \hline\hline
      Filter & Apparent Magnitude \\ 
      \hline
      F090W & > 28.80\\
      F115W & > 28.94\\
      F150W & $28.4^{+0.3}_{-0.2}$ \\
      F200W & $27.81^{+0.12}_{-0.11}$ \\
      F277W & $27.85^{+0.10}_{-0.09}$ \\
      F356W & $27.60^{+0.13}_{-0.12}$\\
      F410M & $27.2^{+0.3}_{-0.2}$\\
      F444W & $27.28^{+0.12}_{-0.11}$\\
      \hline \hline
   \end{tabular}
   \tablefoot{Reported magnitudes are corrected for lensing: $m_{delensed} = m_{observed} + 2.5 \log(1.6) \simeq m_{observed} + 0.51$. Errors are 1$\sigma$, upper limits are 2$\sigma$.\newline
   $^{(\dag)}$ Measured from the centroids of \hg, \hd, \neiii~and \oiiiaur. 
   }
\end{table}

\begin{table*}[htbp]
   \caption{Measured line properties.} 
   \label{tab:lines}      
   \centering          
   \begin{tabular}{c c c c c c}
      \hline\hline
      Line & Observed Wavelength & FWHM & Flux & EW & S/N \\
       & (\micron) & (\AA) & ($10^{-19}$ \ergscm) & (\AA) & \\
      \hline
      \niv & 1.75 & $25^{+7}_{-9}$ & $<2$ & $<3$ & 1.96 \\
      \civblue & 1.83 & / & $<1.2$ & $<2$ & $<2$ \\
      \nev & 4.04 & $60\pm20$ & $<0.72$ & $<5$ & 1.93 \\
      \oiidoublam & 4.40 & $55^{+17}_{-12}$ & $1.8\pm0.4$ & $17\pm4$ & 4.5 \\
      \neiiibluelam & 4.56 & $45\pm4$ & $5.0\pm0.4$ & $48\pm5$ & 12.5 \\
      \heialone\,$\lambda3889$ & 4.59 & $60\pm20$ & $1.2\pm0.4$ & $12\pm0.4$ & 3.0 \\
      \neiiiredlam & 4.68 & $58^{+10}_{-8}$ & $3.0\pm0.5$ & $34\pm6$ & 6.0 \\
      \hd & 4.84 & $37^{+19}_{-12}$ & $1.3^{+0.4}_{-0.3}$ & $16\pm5$ & 3.7 \\
      \hg & 5.12 & $49\pm6$ & $4.6^{+0.6}_{-0.5}$ & $78\pm15$ & 8.4 \\
      \oiiiaur & 5.15 & $49^*$ & $3.4\pm0.4$ & $55\pm11$ & 8.5 \\
      \hline\hline
   \end{tabular}
   \tablefoot{Reported values represent the median and 16th/84th percentiles of the posterior distribution obtained via MCMC sampling. Upper limits are 2$\sigma$. Observed wavelength uncertainties are of the order of 1 \AA. Fluxes are not corrected for lensing. 
   $^{(\ast)}$ Fixed parameters
   }
\end{table*}

\twocolumn
\section{Morphological fit} \label{sec:appendix_morph}
\begin{figure}
\center
 \includegraphics[width=0.49\textwidth]{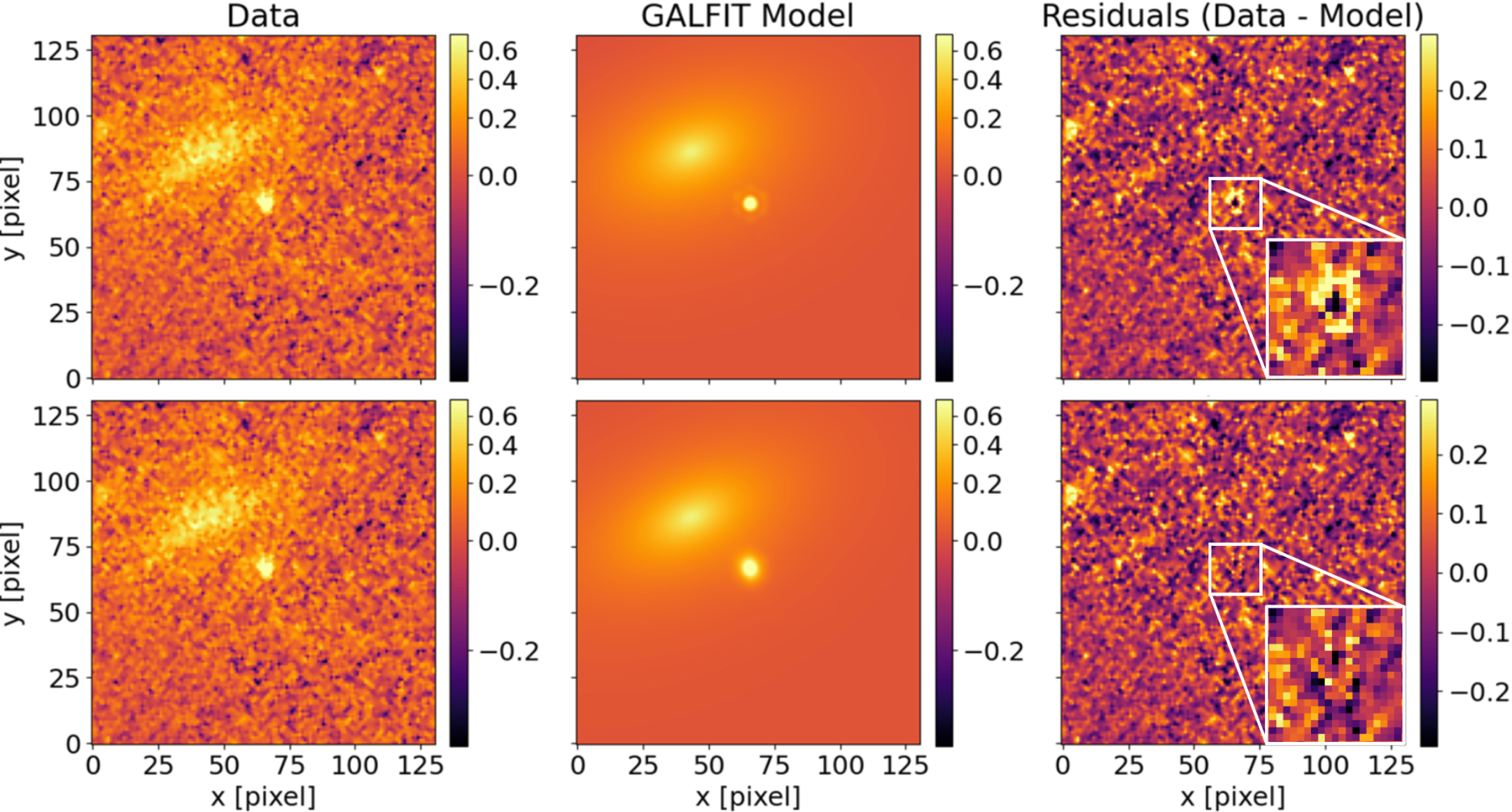}
 \caption{Galfit morphological fits using a three-component model consisting of Janus-11, the foreground galaxy and the sky background for the F200W filter. The columns show, from left to right, the original image, the model and the residuals. The first row corresponds to the fit with a PSF model for Janus-11, where significant residuals are visible; the second row corresponds to the fit with a Sérsic profile for Janus-11.
 }
 \label{fig:galfit}
\end{figure}

We perform morphological fitting using GALFIT \citep{galfit_2}. Because of a $z=0.82^{+0.09}_{-0.10}$ \citep{CanucsDR1_2026} foreground galaxy on the sky-plane, we model both that galaxy and Janus-11, together with a sky model, simultaneously. The model for the foreground galaxy is a Sérsic profile with all parameters free to vary, while for Janus-11 we tested both a PSF model and a Sérsic profile with Sérsic index $n=1$. We find that using a PSF model there are significant residuals in all filters, implying that Janus-11 is spatially resolved, as shown in Figure~\ref{fig:galfit} for the F200W filter, with a (rest-frame UV) magnification-corrected circularized effective radius $r_e = 136 \pm 9\,\mathrm{pc}$. 

We also test the model above by adding a PSF component to Janus-11 with the center fixed to the center of the $n=1$ Sérsic component, finding a 2$\sigma$ upper limit on the PSF total flux contribution of $7\%$ in the F200W filter.

\section{SED fitting with Salim extinction curve} \label{sec:appendix_sed}
\begin{figure*}
\center
 \includegraphics[width=0.99\textwidth]{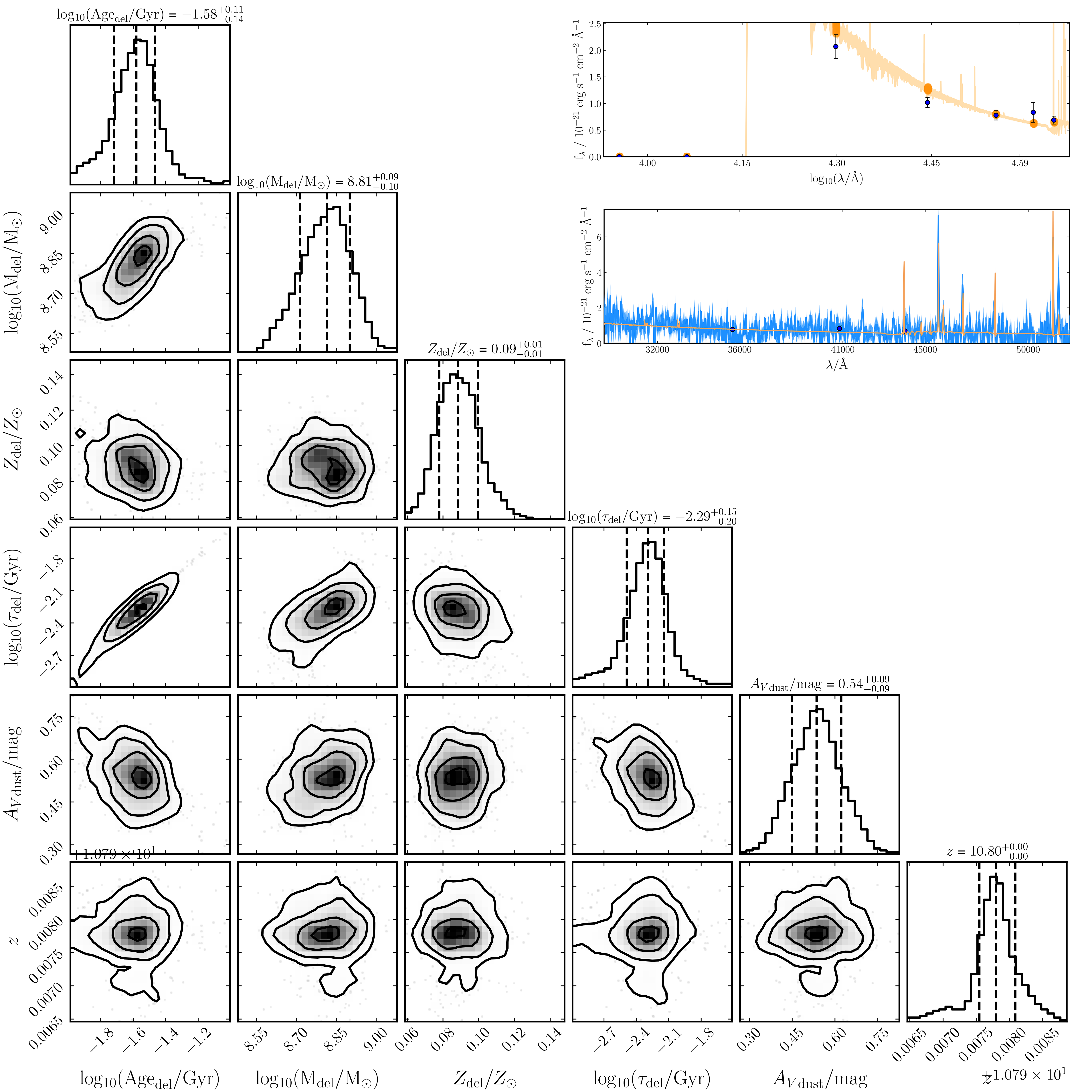}
 \caption{SED fitting posteriors assuming a delayed SFH and a Calzetti dust extinction curve. The upper right plots represent the observed photometry and spectroscopy in blue, and the best-fit model in yellow.
 }
 \label{fig:sed_calzetti}
\end{figure*}

In addition to the SED fitting with the assumptions reported in Section \ref{sec:sedAndMorph} (Figure \ref{fig:sed_calzetti}), we also perform a fit assuming a Salim extinction curve \citep{Salim18}. This curve is based on the Calzetti extinction curve, but is characterized by a parameter $\delta$, which varies the slope, and a parameter B to account for the 2175\AA~bump. Therefore, given the poorly constrained properties of dust obscuration in high-redshift galaxies, this flexibility can be advantageous, as it does not assume a fixed curve, unlike the Calzetti law. The drawback is that to constrain those two additional parameters, more data over a broader wavelength range are required.

\begin{figure*}
\center
 \includegraphics[width=0.99\textwidth]{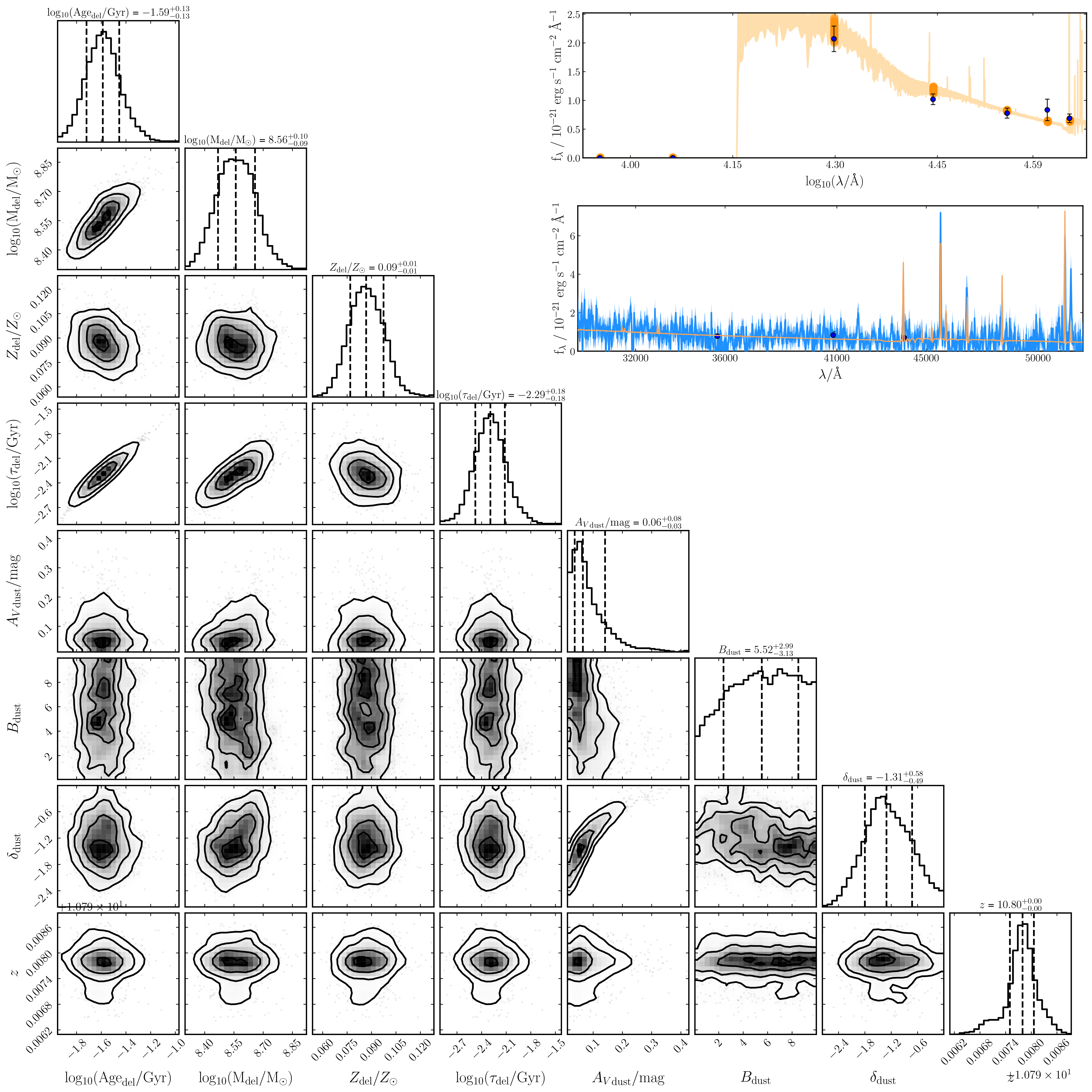}
 \caption{SED fitting posteriors assuming a delayed SFH and a Salim dust extinction curve. The upper right plots represent the observed photometry and spectroscopy in blue, and the best-fit model in yellow.
 }
 \label{fig:sed_salim}
\end{figure*}

For Janus-11, the fit with the Salim extinction curve (Figure \ref{fig:sed_salim}) leads to a smaller stellar mass of log(\mstar/\msun)=$8.50^{+0.10}_{-0.09}$ and a lower SFR$_{10}$ of $5.5^{+1.4}_{-0.8}$\,\sfr than the fit with the Calzetti extinction curve.

\begin{figure*}
\center
 \includegraphics[width=0.95\textwidth]{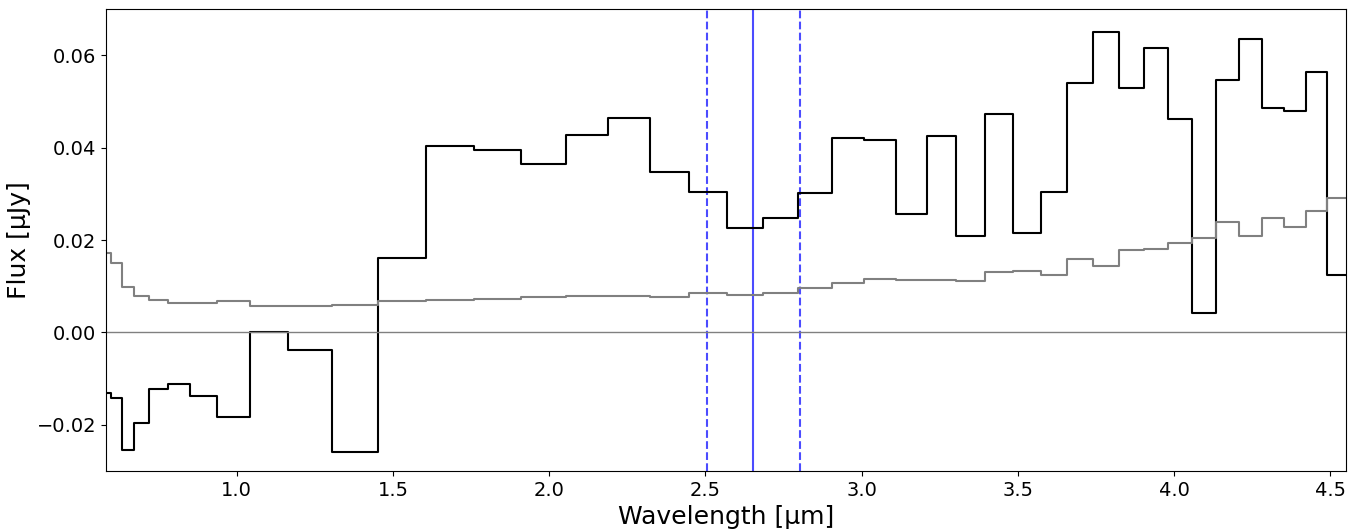}
 \caption{JWST/NIRSpec prism spectrum rebinned with 8 pixels per bin; the grey line represents the 1$\sigma$ uncertainty. The solid blue line is at 2250\AA and the dashed blue lines are at $2250\pm125$\,\AA, where 125 is the expected bump HWHM.
 }
 \label{fig:prism_rebinned}
\end{figure*}

Regarding the dust properties, the estimated $A_V$ value is $0.06^{+0.08}_{-0.03}$, which would classify Janus-11 as a blue monster. However, there is a degeneracy between $A_V$ and the $\delta$ parameter, as can be seen from the corner plot. A negative value of $\delta$, such as the estimated value of $-1.31^{+0.58}_{-0.49}$, leads to a steeper extinction curve in the UV/optical region; therefore, the observed SED could be explained either by a moderate $A_V$ value with a Calzetti-like extinction curve slope or by a relatively low $A_V$ value with a steep slope. In order to break this degeneracy, however, data at redder wavelengths would be required. Also, this $\delta$ value is an extreme value lower than those usually observed even at high redshift and it does not correspond to any of the dust laws observed in the literature \citep{Salmon2016}; for this reason, the prior is often limited to $-1 < \delta < 0.4$, but doing so for Janus-11 results in posteriors that are pushed against the lower prior boundary.\newline
The B parameter is not well constrained by the data, but the posterior distribution seems to disfavor a low value, hinting at the possible presence of the 2175\AA~bump. In order to better constrain this hypothesis, we make use of the NIRSpec prism spectrum from the Canadian NIRISS Unbiased Cluster Survey (CANUCS, \citealt{willott2022}) reduced by DJA \citep{degraaff2025,Heintz2024}. Specifically, we rebin the spectrum using 8-pixel bins, and we find a possible bump signature centered at $\sim$2250\AA, as shown in Figure \ref{fig:prism_rebinned}. This would be consistent with the results of \citet{Witstok2023}, who reported this feature centered at $\sim$2250\AA~for a $z\sim8$ galaxy. However, while both the prism spectrum and the photometry tend to favor the presence of the bump, the data quality is not good enough to exclude the possibility that it is just the result of noise and systematic uncertainties.

\begin{figure*}
\center
 \includegraphics[width=0.95\textwidth]{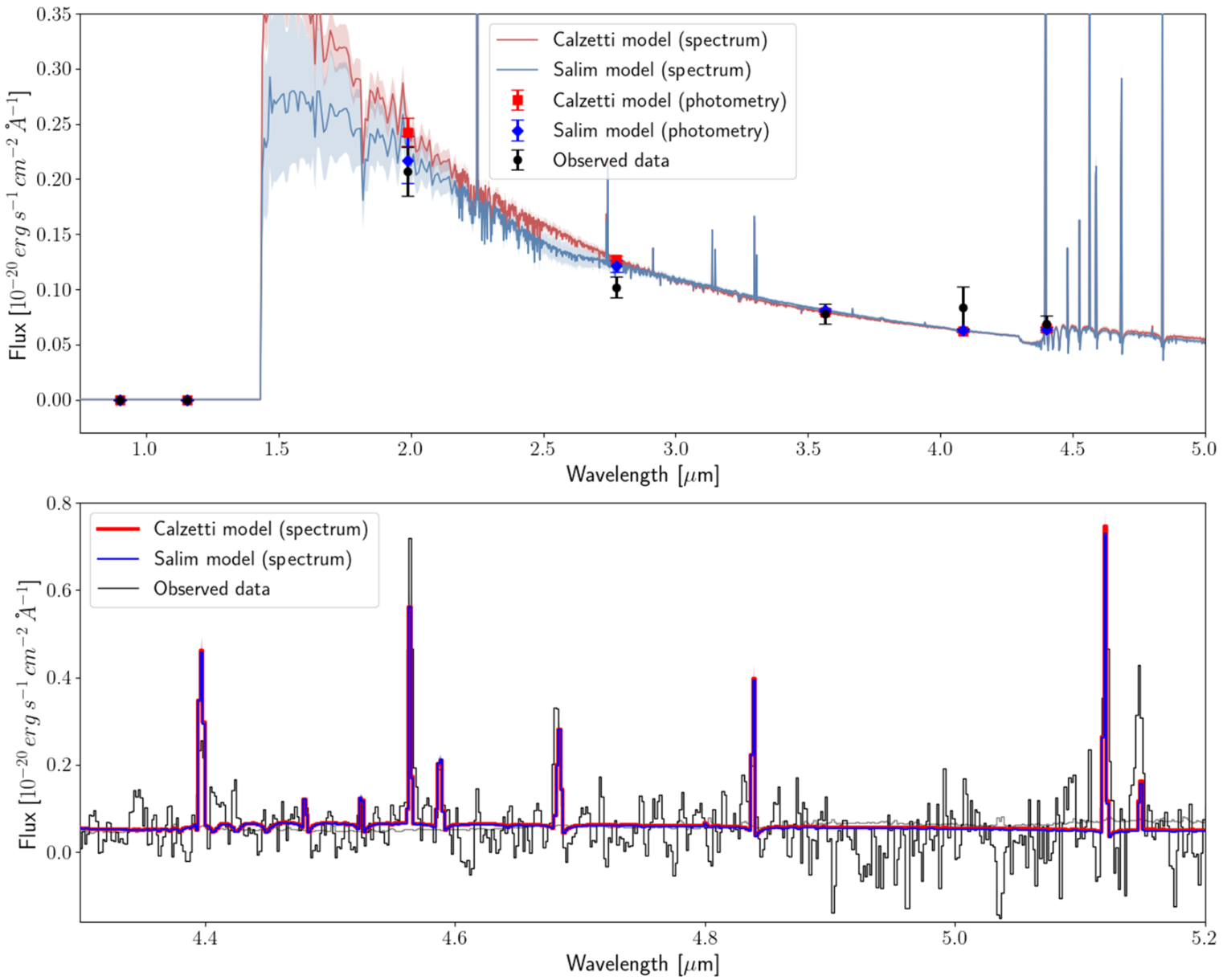}
 \caption{Comparison between the best-fit SED models and the observed data, using both the Calzetti and Salim dust extinction laws. Both the photometric data and the spectrum are corrected for magnification.
 Upper panel: red and blue points show the best-fit photometry for the Calzetti and Salim models respectively, with error bars indicating the 16th and 84th percentile posterior uncertainty; red and blue lines show the best-fit spectra, with shaded regions indicating the 16th and 84th percentile posterior uncertainty; black points show the observed photometric data.
 Lower panel: zoom-in on the spectral region containing prominent emission lines, showing the best-fit model spectra (colours and shaded regions as above) superimposed on the observed spectrum (black) and its 1$\sigma$ noise level (gray).}
 \label{fig:sed_fit_comparison}
\end{figure*}

A comparison between the two models and observed data is shown in Figure \ref{fig:sed_fit_comparison}.


\section{HII-CHI-mistry inferred metallicity and ionization parameter}
\label{sec:HII_CHI_mistry}
We also report the metallicity and ionization parameter computed using the software HII-CHI-mistry version 6.11 \citep{Perez-Montero2014}, which uses a Bayesian-like comparison between line flux ratios with large grids of photoionization models.
From \oiidoublam, \neiiibluelam, \oiiiaur~and \oiiiv~fluxes normalized to the \hb~flux, we find $12 + \log(O/H) = 7.68\pm0.2$ or $Z=0.10^{+0.06}_{-0.04}\,\zsun$ and $\log U=-1.75\pm0.09$; both values are consistent with the ones derived in Sect.~\ref{temp_logu_metal}.
\end{document}